\documentstyle[aps,prl]{revtex}  %% Journal Style 
\begin{document}
\draft

%% Include following two lines for Journal Style

\twocolumn[\hsize\textwidth\columnwidth\hsize  %**Journal
\csname @twocolumnfalse\endcsname              %**Journal

\title{\bf  Electroproduction of the $d^*$ dibaryon } 

\author { Chun Wa Wong}

\address{
Department of Physics and Astronomy, University of California, 
Los Angeles, CA 90095-1547}

\date{January 31, 2000}

\maketitle

\begin{abstract}
The unpolarized cross section for the electroproduction of 
the isoscalar  $J^\pi = 3^+$ di-delta dibaryon $d^*$ is 
calculated for deuteron target using a simple picture of 
elastic electron-baryon scattering from the $\Delta \Delta (^7D_1)$  
and the $NN (^3S_1)$ components of the deuteron. The calculated 
differential cross section at the electron lab energy of 1 GeV 
has the value of about 0.24 (0.05) nb/sr at the lab angle of 
10$^\circ$ (30$^\circ$) for the Bonn B potential when the dibaryon 
mass is taken to be 2.1 GeV. The cross section decreases rapidly 
with increasing dibaryon mass. A large calculated width of 40 MeV 
for $d^*(\Delta\Delta\,^7S_3)$ combined with a small experimental 
upper bound of 0.08 MeV for the $d^*$ decay width appears to have excluded any low-mass $d^*$ model containing a significant 
admixture of the $\Delta\Delta (^7S_3)$ configuration.
\end{abstract}


%%  Include the following line for Journal style
     ]  %**Journal

\narrowtext

\section{Introduction}
\label{sec:level1}

Given the availability of excellent electron beams, the electroproduction of dibaryon resonances \cite{Wong78} 
represents a promising way to look for these long-sought 
objects \cite{Jaf77,LISS95}. In assessing 
the practical prospects of such experiments, one needs to begin 
with rough estimates of the electroproduction cross sections. This 
has to be done separately for each dibaryon candidate of interest, 
since each candidate may involve unique theoretical issues.

One candidate dibaryon that has been discussed recently is the 
di-delta $d^*$ of quantum numbers $J^\pi T = 3^+0$ 
\cite{Gold89,Wang92,Gold95,Wong98}. This dibaryon is a six-quark 
state which 
at large inter-baryon separations may be visualized as a pair 
of $\Delta$'s. Preliminary studies of its electroproduction 
from deuteron targets have recently been made by Qing 
\cite{Qing97} and by Sun \cite{Sun97}, both of Nanjing University.
They find that the Kroll-Ruderman $\gamma d \rightarrow 
\Delta\Delta$ production process does not contribute because of 
the special isospin structure of $d$ and $d^*$. They concentrate 
instead on another process important in the electroproduction 
of two pions from proton \cite{Gom94} and deuteron \cite{Gom95}
 based on the electroexcitation of a nucleon isobar $N^*(1520)$ in 
the intermediate state, namely
$\gamma d \rightarrow NN*(1520) \rightarrow \Delta\Delta$. This mechanism can be interpreted as the production of $d^*$ through 
its $NN^*$ component.

However, these preliminary results for the calculated 
differential cross sections turn out to be very small at small 
momentum transfers (say $< 500$ MeV/$c$), 
almost two orders of magnitude smaller than a simple estimate 
made by me using a simple picture of inelastic scattering from 
the $\Delta\Delta (^7D_1)$ of the deuteron, as described in 
diagram a of Fig. \ref{figFeyn}. These results seem to suggest 
that  $d^*$ is more easily produced from the deuteron at these 
small momentum transfers through elastic electron-baryon (eB) 
scatterings rather than inelastic excitations of the struck baryon.

It is worth noting in connection with diagram \ref{figFeyn}a that 
if the $d^*$ is visualized as a $^7S_3$ didelta, there is no 
production from the deuteron $\Delta \Delta (^3S_1)$ component in 
the lowest order. This is because the 
electromagnetic operators are of rank only 0 or 1 in intrinsic 
spin. Hence electroproduction in this simple lowest-order 
picture is possible only from the $\Delta \Delta (^7D_1)$ 
component of the deuteron.

In addition to diagram \ref{figFeyn}a, or equivalently its 
perturbative part diagram \ref{figFeyn}b, $d^*$ can also be 
electroproduced through the $NN(^3D_3-^3G_3)$ component of 
$d^*$, as described in diagram \ref{figFeyn}c. Lomon has 
looked for a $d^*$ resonance in these 
channels by studying the energy dependence of their $NN$ phase
 parameters \cite{Lom97,Lom99}. These $NN(^3D_3-^3G_3)$ 
channels are in fact the dominant decay channels of $d^*$ \cite{Wong98,Wong98a,Wong99}. He has found that 
these phase parameters are consistent with the absence of any 
dibaryon resonance with a width exceeding 1 (2) MeV near 2.1 
(2.25) GeV. 

A much stricter upper limit on the width of an np dibaryon 
resonance has been obtained experimentally some time ago by 
Lisowski {\it et al.} \cite{Lis82}. They measured the total 
$np$ cross section at c.m. energies 2.00 to 2.23 GeV at high 
energy resolution, namely about 1.4 MeV at 2.11 GeV. No evidence 
was seen for narrow resonances with areas greater than 5 mb-MeV. 
At 2.1 GeV where the 
$np$ $I=0$ total cross section is 33 mb, a totally elastic $^3D_3$ resonance would have a maximum unitarity-limited cross section 
of $\sigma_{\rm max} = 39$ mb. Its energy-integrated area is $(\pi/2)\sigma_{\rm max}\Gamma$ for a pure Breit-Wigner shape, 
where $\Gamma$ is its total decay width. The assumption of a 
pure Breit-Wigner resonance should be a very good one, because the 
$NN$ $^3D_3$ phase shift at the total energy of 2.1 
GeV is only 4$^\circ$ \cite{Lom97}. The experimental bound of 
5 mb-MeV on the integrated area translates to an upper bound of 
only 0.08 MeV for the width of any resonance that has escaped 
detection in this experiment. 

The theoretical decay width of $d^*$ has been estimated in
 \cite{Wong98}. If the $d^*$ is taken to be a bound system of 
two finite-sized $\Delta$'s, describable by a nonrelativistic 
quark model, the decay has been estimated to be about 10 MeV 
when its mass is 2.1 GeV. The result is, however, sensitive to 
both dynamical input and rescattering corrections; it could be 
as large as 40 MeV for a realistic 
potential treated perturbatively without any rescatterinfg 
correction, as I shall show in more detail in this paper. 

It has been argued in \cite{Wong98} that if the dibaryon is made 
up of six delocalized quarks \cite{Gold89,Wang92,Gold95} instead 
of two separated $\Delta$'s, 
its decay width could well be smaller by an order of magnitude, 
say 1 MeV. This suggestions remains to be confirmed by a 
detailed calculation. A reduction to 0.1 MeV would be much 
harder to realize. Furthermore, if the estimated decay width has 
to be reduced for any reason, its estimated production cross 
sections, including the electroproduction cross section estimated 
here using a didelta model, must also be reduced correspondingly. 
 
Thus the case for a $d^*$ resonance near 2.1 GeV does not appear
 promising. However, the $d^*$ could have a mass higher than 
expected if the dynamnics is different from that described by 
the delocalization and color-screening model. Hence it is still 
of interest to study the electroproduction cross section for 
$d^*$, should such a resonance exist. 

It is obvious that in addition to diagram \ref{figFeyn}a or b, 
one should also included the process 
shown in diagram \ref{figFeyn}c, where the $d^*$ is 
produced through the $NN(^3D_3)$ channel, the $NN(^3G_3)$ 
contribution having been ignored in this lowet-order picture.

Although these two processes alone do not add up to a 
quantitative description of the electroproduction, they are 
of sufficient interest to justify the detailed report given 
here, including the contributions of convective and 
magnetization currents. The picture is necessarily very 
rough, because of the neglect of many components and processes. 
However, uncertainties about the mass and structure of $d^*$ and 
about short-distance nuclear dynamics have discouraged me from 
undertaking a more ambitious calculation at this time.

The paper is organized as follows: The notation used is defined 
in Sec. II where brief comments relevant to the present 
calculation are made. Sec. III shows how the calculation is 
done for diagram \ref{figFeyn}a, when the needed 
$\Delta \Delta (^7D_1)$ component of the deuteron is already 
available. The contributions from different reduced matrix 
elements (RME's) are briefly discussed. 

  Sec. IV shows that the inclusion of the $d^*(NN\,^3D_3)$ 
contribution via the perturbative process Fig. \ref{figFeyn}c 
can be broken 
down into three steps: (a) a perturbative evaluation of the
$d^*(NN\,^3D_3)$ wave function, (b) a calculation of the $d^*$ 
decay width for the same input dynamics (taken to be the 
28-channel Argonne potential \cite{Wir84} and the Bonn B 
potential \cite{Mac89}), and finally (c) the evaluation of 
the electroproduction cross section itself. 

  Calculational results are presented in Sec. V, where the 
production amplitude from diagram \ref{figFeyn}c is found to be 
greater than that from diagram \ref{figFeyn}a by a factor of 
2-4 at certain angles or momentum transfers. At a lab angle of 
$30^\circ$ and an electron lab energy of 1 GeV, I find a 
differential production cross section of about 0.05 nb/sr when 
integrated over the energy loss. 

  My calculated cross section is smaller by two orders of 
magnitude from the preliminary value of about 10 nb/sr obtained 
recently by the Nanjing group (to be called QSW) 
\cite{Qing99} using a different method of calculation. Part of 
the discrepancy, accounting for a factor of 5, comes from the 
fact that QSW has included only $\pi$ exchange but not $\rho$ 
exchange. The remaining discrepancy, of more than an order of 
magnitude, must be due to differences in calculational methods 
used. For example, I use a principal-value Green function for 
two nucleons in the $d^* (NN\, ^3D_3)$ wave function, whereas 
QSW use an outgoing-wave boundary condition. To help in 
disentangling the discrepancy in the future, I have included 
many details in the present paper.

The decay width calculated here for the $\Delta\Delta (^7S_3)$ 
model of $d^*$ is about 40 MeV at $m^* = 2.1$ GeV. When used with 
the experimental upper bound of only 0.08 MeV of any $I=0, J=3$ 
np resonance \cite{Lis82}, the calculated width makes it very 
unlikely that any undetected $d^*$ has a significant probablity 
of the $\Delta\Delta (^7S_3)$ configuration.

\section{ Electroproduction cross section for $ed \rightarrow ed^*$ }

Consider the scattering of a relativistic electron beam of lab 
energy $\epsilon$ to the lab angle $\theta$ by a target that goes 
from an initial state $i$ to a final state $f$ as the result of 
the scattering. The differential cross section in the lab 
(after integrating over the energy transfer) is known to have the 
form \cite{deF66}

\begin{equation}
{{\rm d}\sigma_{\rm fi} \over {\rm d}\Omega} = 
\sigma_M f_{\rm rec}^{-1}[v_L T_{\rm fi} ^L + 
v_T (T_{\rm fi}^{\rm el} + T_{\rm fi}^{\rm mag}) ]\,, 
\label{eq:dcs}
\end{equation}
where

\begin{eqnarray}
\sigma_M  =  \left [ \alpha \cos(\theta /2)\over 2\epsilon 
\sin^2(\theta /2) \right ]^2
\label{Mott}
\end{eqnarray}
is the Mott differential cross section, $\alpha$ is the fine 
structure constant,

\begin{eqnarray}
f_{\rm rec} = 1 + { 2\epsilon \sin^2(\theta /2)\over M_{\rm target} }
\label{f-rec}
\end{eqnarray}
is a target recoil factor,

\begin{eqnarray}
v_L =  \left ( Q^2\over q^2 \right )^2,  \quad 
v_T  =  - {1\over 2} \left ( Q^2\over q^2 \right ) + 
\tan^2 \left( \theta\over 2 \right)
\label{vLvT}
\end{eqnarray}
are the electron kinematical factors that depend on its 
four-momentum transfer $Q = K - K' = (\omega, {\bf q})$. Here 
$K = (\epsilon, {\bf k})$ is the four-momentum of the electron 
in the lab before the scattering, and $K'$ its four-momentum after 
the scattering. The notation is that of  \cite{Don86}, with 
$Q^2 \le 0$. 

  The target factors are

\begin{eqnarray}
T_{\rm fi}^L  =  {4\pi\over 2J_i + 1} \sum_{J=0}^\infty  
\vert \langle J_f \Vert {\hat M}_J(q)\Vert J_i \rangle \vert^2 \, ;  
\nonumber \\
T_{\rm fi}^\alpha  =  {4\pi\over 2J_i + 1} \sum_{J=1}^\infty 
\vert \langle J_f \Vert {\hat T}_J^\alpha (q)\Vert J_i \rangle 
\vert^2 \, ,
\label{targetT}
\end{eqnarray}
where $\alpha =$ el or mag. The nuclear reduced matrix elements (RME)
$\langle J_f \Vert {\cal O} \Vert J_i \rangle $ that appear are all 
dimensionless quantities. The operators $\cal O$ are the longitudinal 
Coulomb, and the transverse electric and magnetic, multipole operators
that have the following simple forms in momentum space \cite{Qing97}

\begin{eqnarray}
{\hat M}_J(q) = {(-i)^J\over 4\pi} \int & 
d\Omega_{\bf q} Y_J(\Omega_{\bf q}) \rho({\bf q}) \, , 
\quad \quad \nonumber \\
{\hat T}_J^{\rm el}(q)  =  -\sqrt{8\pi} \sum_\kappa & 
\langle \kappa \Vert Y_J \Vert 1 \rangle 
\left \{ {\begin{array}{ccc}
                       1 & 1 & 1 \\
                       J  & J & \kappa
          \end{array}} \right \}  \nonumber \\
\times {(-i)^J\over 4\pi} \int  & 
d\Omega_{\bf q} [Y_\kappa (\Omega_{\bf q}) \otimes 
{\hat {\bf J}}({\bf q})]^J \,, \nonumber \\
{\hat T}_J^{\rm mag}(q) = {(-i)^J\over 4\pi} \int & 
d\Omega_{\bf q} [Y_J(\Omega_{\bf q}) \otimes 
{\hat {\bf J}}({\bf q})]^J \,.
\label{multipoles}
\end{eqnarray}
Here $Y_J$ is a spherical harmonic, $ \rho({\bf q})$ and 
${\hat {\bf J}}({\bf q})$ are the baryon charge and current 
density operators. 

  The current density operator can be separated into convective 
and magnetization terms:

\begin{eqnarray}
{\hat {\bf J}}({\bf q}) =  {\hat {\bf J}}_{\rm c} ({\bf q}) + 
{\hat {\bf J}}_{\rm S} ({\bf q}) \, . 
\label{currents}
\end{eqnarray}
The spin magnetization term ${\hat {\bf J}}_{\rm S} = 
\mbox{\boldmath $\nabla$} \times 
{\hat {\mbox{\boldmath $\mu$}}}_{\rm S}$ originates from the 
dibaryon isoscalar magnetic moment operator 
${\hat {\mbox{\boldmath $\mu$}}}_{\rm S} =  
\mu_0 \mu_{\rm s}{\hat {\bf S}}$. Here $\mu_0$ is the nuclear 
magneton, ${\hat {\bf S}}$ is the total spin 
angular momentum operator of the nucleus. In this notation, the 
nucleon magnetic-moment operator is written as 
${\hat {\mbox{\boldmath $\mu$}}}(N) =  
\mu_0 [\mu_{\rm s}(N){\hat {\bf S}} + \mu_{\rm v}(N)
\hat {\mbox{\boldmath $\tau$}}_3{\hat {\bf S}}]$, while the 
isoscalar part of the operator for $\Delta$ is 
$\mu_0 \mu_{\rm s}(\Delta){\hat {\bf S}}$.
The isoscalar baryon magnetic-moment parameters used are 
the experimental value

\begin{eqnarray}
\mu_{\rm s} (N) =  \mu_{\rm p} + \mu_{\rm n} = 0.880
\label{Nmoment}
\end{eqnarray}
for the nucleon, and the theoretical value from the 
non-relativistic quark model

\begin{eqnarray}
\mu_{\rm s} (\Delta) =  \mu_{\rm v}(N)/5 = 0.94
\label{Delmoment}
\end{eqnarray}
for the $\Delta$. This quark model actually gives the same value 
for both baryon isoscalar magnetic moments.  I prefer to use 
the slightly larger value shown for $\mu_{\rm s}(\Delta)$ obtained 
from the experimental nucleon isovector magnetic moment of 
$\mu_{\rm v} (N) = \mu_{\rm p} - \mu_{rm n} = 4.71$.

As is known\cite{Uber71}, the orbital magnetization term is 
already contained in ${\hat {\bf J}}_{\rm c}$, and does not have 
to appear explicitly. 
 
For specific components in the initital and final nuclear states, 
only a few terms are allowed in the multipole sum shown in 
Eq. (\ref{targetT}) by the triangle rule for angular momenta.   
For the dibaryon components included in our calculations, the 
relative BB orbital angular momenta of initial and final nuclear 
states  $(L_i, L_f)$ are either (0,2) or (2,0). Then only the 
$J=2$ multipole term appears for the Coulomb target factor.

The situation for the transverse RME's is slightly more 
complicated. It is controlled by the spatial RME

\begin{eqnarray}
\langle L'=0\Vert \int d\Omega_{\bf q} [Y_\kappa (\Omega_{\bf q}) 
\otimes \bar {\bf p}]^\lambda \Vert L=2 \rangle \nonumber \\
=  \delta_{\lambda 2} \sum_{\kappa = 1,3} f_\kappa(q) \, ,
\label{spatialRME}
\end{eqnarray}
where $ \bar {\bf p} \equiv {\bf p} + ({\bf q}/4)$ is the mean 
of the initial relative BB momentum ${\bf p}$ and its final 
value ${\bf p} + ({\bf q}/2)$ after the absorption of a virtual 
photon of momentum ${\bf q}$. Part of the photon momentum, 
${\bf q}/2$, goes into the recoil of the dibaryon. The 
functions $f_\kappa(q)$ are not needed at this point, and will 
not be given. The important features are that the operator must 
be a spatial quadrupole operator, and that $\kappa$ can only be 
1 or 3.

Eq. (\ref{spatialRME}) has the consequence that the only 
operators contributing to the transvers RME for the dibaryon wave 
functions used in the present study are (a) the convective and 
orbital magnetization current terms in ${\hat T}_{J=2}^{\rm el}$, 
(b) the spin magnetization current term in ${\hat T}_J^{\rm el}$,
with $J=$2 and 3, and (c) the spin magnetization current term in 
${\hat T}_{J=3}^{\rm mag}$.

\section{ Contributions of the deuteron 
$\Delta \Delta (^7D_1)$ component}

The $d^*$ dibaryon will be treated in this paper as a pure $\Delta\Delta(^7S_3)$ state, using the two-centered Gaussian 
wave function parametrized in \cite{Wong99} as a sum of Gaussians. 
This wave function will be specified with other wave functions 
used in this paper in the Appendix. 

The inelastic production of $d^*$ 
from the deuteron $\Delta\Delta(^7D_1)$ component in the initial 
state by elastic $e\Delta$ scattering 
is described by diagram \ref{figFeyn}a. The Argonne-28 (A28) 
deuteron $\Delta\Delta(^7D_1)$ wave functions \cite{Wir99} are 
used, expanded in harmonic-oscillator wave functions (HOWF). 

However, it is sufficient to give explicit expressions for the RME's 
appearing in the differential production cross section only for 
single-term wave functions such as the single Gaussian 

\begin{eqnarray}
\psi_{d^*}({\bf p}) = N_0^* e^{-p^2/2\beta^{*2}}
\label{Gaussian}
\end{eqnarray}
for $d^*$. Here  

\begin{eqnarray}
N_0^* = (\pi\beta^{*2})^{-3/4} \, .
\label{n0star}
\end{eqnarray}
For the deuteron $\Delta\Delta(^7D_1)$ component, a one-term form 
of the HOWF is

\begin{eqnarray}
\psi_{d}({\bf p},\Delta\Delta, ^7D_1) = \sqrt{P_{\Delta\Delta 7}} N_2 
{\rm e}^{-p^2/2\beta^2} {\cal Y}_{\rm 2m}({\bf p})\, ,
\label{d7d1-wf}
\end{eqnarray}
where $P_{\Delta\Delta 7}$ is the $\Delta\Delta(^7D_1)$-state 
probability of the deuteron, 

\begin{eqnarray}
N_2 = \left (16\pi\over 15\beta^4 \right )^{1/2}(\pi\beta^2)^{-3/4}\, ,
\label{n2}
\end{eqnarray}
and ${\cal Y}_{\rm 2m}({\bf p}) = p^2 Y_{\rm 2m}(\hat {\bf p})$ 
is a solid spherical harmonic. Then the Coulomb RME is

\begin{eqnarray}
\langle d^*\Vert {\hat M}_2(q) \Vert d_7\rangle = 
\sqrt{3/5} \, \langle d^*,S\Vert {\hat M}_2(q)\Vert d_7, D\rangle 
\nonumber \\
=  - {\sqrt{3}\over 4\pi} (\lambda q)^2 f(q)\, ,
\label{CoulRME}
\end{eqnarray}
where $d_7$ stands for the deuteron $\Delta\Delta(^7D_1)$ component;
\begin{eqnarray}
\lambda = { a^*\over 2(a^*+b) } \, ;
\label{ablambda}
\end{eqnarray}
\begin{eqnarray}
f(q) = \sqrt{P_{\Delta\Delta}} N_0^* N_2\left ( \pi\over 
a^*+b \right )^{3/2}  
{\rm e}^{-cq^2/2} \, ;
\label{fq}
\end{eqnarray}
\begin{eqnarray}
a^* = (2\beta^{*2})^{-1}, \quad b = (2\beta^2)^{-1}; \nonumber \\ 
\quad c = \lambda b + \alpha_0; 
\quad \alpha_0 = r_p^2/3 = 0.12 fm^2 \, .
\label{abc}
\end{eqnarray}
The $\alpha_0$ term takes care of the baryon form factor at 
the $\gamma$-baryon vertex. If the $\Delta$ is assumed for 
simplicity to have the same size as the nucleon, the same 
Gaussian form factor appears in all terms of the production 
amplitude in the nonrelativistic quark model. Additional 
baryon form factors that should appear are already included 
in the BB interactions themselves.

   The transverse RMS's can be separated into convective, and 
spin magnetization terms:

\begin{eqnarray}
\langle d^*\Vert {\hat T}_{\rm c2}^{\rm el}(q) \Vert d_7\rangle  
& = & - {\sqrt{2}\over 4\pi} 
{1\over M_{\rm B}} g_\lambda(q)f(q)\, ,
\nonumber \\
\langle d^*\Vert {\hat T}_{\mu 2}^{\rm el}(q) \Vert d_7\rangle  
& = & - (\mu_{\rm s}(B) \mu_0/ \pi) f(q) \nonumber \\
&& \times [g_\lambda (q) - g_\nu (q)] 
\, , \label{elRME}
\end{eqnarray}
where $B = \Delta$ in $M_{\rm B}$ and $\mu_{\rm s}(B)$, and

\begin{eqnarray}
g_\lambda(q) = {3\lambda q\over 2(a^*+b)}, \quad g_\nu (q) = 
\lambda^2 \nu q^3, \quad 
\nu = {1\over 4} - \lambda \, ;
\label{g}
\end{eqnarray}
\begin{eqnarray}
\langle d^*\Vert {\hat T}_{\ c3}^{\rm mag}(q) \Vert d_7\rangle  
= \langle d^*\Vert {\hat T}_{\rm L3}^{\rm mag}(q) \Vert d_7\rangle  
= 0 \, , \nonumber \\
\langle d^*\Vert {\hat T}_{\rm S3}^{\rm mag}(q) \Vert d_7\rangle  
= - \sqrt{8/7}\, {\mu_0\over 4\pi} \mu_{\rm s}(B) g_\nu (q) f(q)
\, .
\label{magRME}
\end{eqnarray}

  I use a 3-term approximation to the deuteron 
$\Delta\Delta(^7D_1)$ wave function of the Argonne 28-channel 
potential \cite{Wir84,Wir99}, denoted below as potential A28.
The approximate wave function used is given in the Appendix.

  It is convenient to present the calculated results not as 
angular distributions which contain strong dependence on the 
electron energy, but as the effective $T$ factor

\begin{equation}
T_{\rm eff} = T_{\rm fi} ^L + (v_T/v_L) (T_{\rm fi}^{\rm el} + T_{\rm fi}^{\rm mag})\,. 
\label{effFF}
\end{equation}
The Coulomb $T$ factor $T^L$ is a function only of the 
three-momentum transfer $q$, and is energy-independent in our 
model. The transverse terms contain the kinetmatical factor 
$v_T/v_L$, but the energy dependence from the angle-dependent 
term is small except at low electron energies. 

The weak energy dependence of these effective $T$ factors is 
explicitly illustrated in Fig. \ref{figEffT1a} 
for a $d^*$ mass $m^*$ of 2100 MeV and for three different 
electron energies. The A28 $d(\Delta\Delta\, ^7D_1)$ wave 
function is used. The three-momentum  transfer has the 
energy-independent minimal value of 
$q_{\rm min} \approx 238$ MeV/$c$ at this value of $m^*$. 
The momentum range shown corresponds to an angular range of 
about $90^\circ$ ($15^\circ$, $2^\circ$) for 1 GeV 
(4 GeV, 27.5 GeV) electrons.

  In the transverse electric $T$ factor, the amplitude from 
the magnetization current is a factor of 2.5 or 3 larger than 
that from the convective current. This means that the electric 
$T$ factor is larger by roughly an order of magnitude when the 
contribution from the magnetization current is included. 

On the other hand, the transverse magnetic $T$ factor is very 
small at small angles, but has become 1/3 as large as the 
transverse electric $T$ factor at $q \approx 1$ GeV/$c$.

The calculated cross section for diagram \ref{figFeyn}a is 
proportional to the probability $P_{\Delta\Delta 7}$ of the 
$\Delta\Delta (^7D_1)$ component of the deuteron, other things 
being equal. Besides the A28 
model, the Argonne group has constructed a weaker model with 
$P_{\Delta\Delta 7} = 0.23 \%$. The coupled-channel models E and F 
constructed by Dymarz and Khanna \cite{Dym90} have $P_{\Delta\Delta 7} 
\approx 0.1-0.4 \%$, and a total $P_{\Delta\Delta}\approx 0.4-0.5 \%$. 
The relativistic field-theory model studied by Ivanov {\it et al.}
 \cite{Iva99} gives a total $P_{\Delta\Delta}$ of only 0.08\%. 

  These theoretical estimates are consistent with the best 
experimental information on the total $P_{\Delta\Delta}$, namely 
that it does not exceed 0.4\% at 90\% CL from the null result of 
a bubble chamber search for the spectator ${\Delta^{++}}$ in a 
$\nu d$ knockout reaction \cite{All86}.

\section{ Inclusion of the $NN (^3D_3)$ component of $d^*$ }

Although diagram \ref{figFeyn}c has a structure rather similar 
to that for diagram \ref{figFeyn}b, its production amplitudes 
are much harder to calculate because of three complications: 
(1) The wave function in the minor 
component $d^*(NN\,^3D_3)$ needed in the calculation is not to 
our best knowledge readily available in the literature. It has 
to be evaluated {\it ab initio}. 
(2) This wave function is very sensitive to the spreading width 
$\Gamma$ of $d^*(\Delta\Delta\,^3S_3)$ into $NN$ channels. Although
the production cross section turns out to be only mildly dependent 
on $\Gamma$, it is desirable to use the width consistent with the 
assumed input dynamics in the calculation. This too has to be 
evaluated. (3) The wave function has sharp kinks near $p_0$ where 
the nucleon energy $E_N(p_0)$ has the value $m^*/2$. We find it 
simpler to perform a one-dimensional integration for the 
production amplitudes instead of using the harmonic-oscillator 
expansion described in the last section. 

The calculation is thus broken up into three major steps in 
order to handle the three complications listed above.

\subsection{ Perturbative treatment of wave-function components }

  Our first concern is to estimate the accuracy of a 
perturbative treatment of $d^*(NN\,^3D_3)$. This is done by 
first examining a similar perturbative generation of the 
$d(\Delta\Delta\,^7D_1)$ from $d(NN\,^3S_1)$ using the 
Argonne-28 potential from \cite{Wir84}:

\begin{eqnarray}
\psi(\Delta\Delta\,^7D_1,p) \equiv \langle \Delta\Delta\,^7D_1,p
\vert d \rangle \nonumber \\
\approx \langle \Delta\Delta\,^7D_1,p\vert {1\over \Delta H} V
\vert d(NN\,^3S_1) \rangle \nonumber \\
= {1\over m_d - 2E_\Delta(p)} {1\over (2\pi)^3} 
\int  \langle \Delta\Delta ^7D_1\vert {\cal O}_{18}(\hat {\bf q})
\vert NN\, ^3S_1 \rangle  \nonumber \\
 \times v_{18}(q) \langle {\bf p}+{\bf q}\vert d(NN\, ^3S_1) 
\rangle d^3{\bf q} \, . 
\label{DD7D1}
\end{eqnarray}
Here $\Delta H = m_d - H_0$, with $H_0$ the unperturbed 
Hamiltonian and $m_d$ the deuteron mass, $E_\Delta(p)$ is the 
nonrelativistic $\Delta$
energy, ${\cal O}_{18}(\hat {\bf q})$ is the 
$NN \leftrightarrow \Delta\Delta$ tensor-force operator 
$S_{12}^{\rm III}(\hat {\bf q}) {\bf T}_1\cdot {\bf T}_2$ in the 
A28 notation with ${\bf T}_i$ the $N \leftrightarrow \Delta$ 
isospin operator. The subscript 18 refers to channel 18 of A28.
The momentum-space potential is 

\begin{eqnarray}
v_{18}(q) = - v_{\pi 0}^{\rm III} {4\pi\over \mu^3}
{q^2\over q^2 + \mu^2} f(q) \, ,
\label{v18}
\end{eqnarray}
where $\mu = 138$ MeV is the pion mass, 
$ v_{\pi 0}^{\rm III} = 14.91$ MeV, 
and $f(q)$ is a numerically generated cutoff function that 
decreases from 1 at $q=0$ to 0 at $q = \infty$. 

  Equation (\ref{DD7D1}) can be simplified to a one-dimensional 
integral if the deuteron $NN(^3S_1)$ wave function is expanded 
as a sum of Gaussians. It is sufficient to give the final 
expression for a single Gaussian wave function of the form of 
Eq. (\ref {Gaussian}), but with falloff parameter $\beta$ and an 
associated normalization constant $N_0$ if the S-state probability 
were 100\%. Then

\begin{eqnarray}
\psi_7(p) \equiv \psi(\Delta\Delta\,^7D_1,p) \approx
{h(p)\over m_d - 2E_\Delta(p)} \, ,
\label{DD7D1final}
\end{eqnarray}
where the subscript 7 refers to the spin degeneracy,

\begin{eqnarray}
h(p) = \sqrt{28\over 5\pi^3} N_0 
{\rm e}^{-p^2/2\beta^2} \int v_{18}(q) {\rm e}^{-q^2/2\beta^2} 
\nonumber \\ \times j_2(ix) q^2 dq \, ,
\label{hp}
\end{eqnarray}
and $x=pq/\beta^2$.
 
 The actual calculation is made with a three-term fit to the A28
deuteron $NN(^3S_1)$ wave function \cite{Wir84,Wir99} given in 
the Appendix.
The resulting deuteron $\Delta\Delta(^7D_1)$ wave function 
calculated from Eq. (\ref{DD7D1final}) (long dashed curve) is 
compared in Fig. \ref{fig7d1wf} with the actual A28 wave 
function (solid curve) in the fitted form given 
by Eqs. (\ref{fittedwf})-(\ref{amp}). The wave function is so 
defined that $\int \psi_7^2(p) p^2 dp$ gives the fractional 
normalization $P_7$ in this  $\Delta\Delta(^7D_1)$ state. The 
calculated value in the perturbative approximation 
is $P_7 = 0.65\%$ compared to the exact value of 0.419\%.

  One can see from Fig. \ref{fig7d1wf} that the inaccuracy in the 
perturbative treatment comes from the overestimate of the wave 
function at all momenta larger than $\approx 1.3\mu$, where 
$\mu = 138.0$ MeV is the average pion mass used in these 
calculations. The error 
seems to arise primarily from the neglect of a short-range 
repulsive central potential in the $\Delta\Delta(^7D_1)$ channel. 
Such a repulsive potential would have reduced the wave function 
at small inter-baryon distances. The size of the error in our 
perturbative treatment in this $^7D_1$ state seems to be much 
greater than the results found by \cite{Dym90}. We shall see 
that there are much greater uncertainties elsewhere in the 
present calculation. Consequently, I shall consider the accuracy 
of the perturbative treatment adequate for the present 
qualitative study.

  The same perturbative method is now used to obtain the 
$NN(^3D_3)$ component of $d^*$. An added complication appears 
because the $\Delta\Delta(^7S_3)$ component of $d^*$ is a 
bound state embedded in the $NN$ continuum. The calculation will 
require the use of a spreading width into the $NN$ continuum 
(which is essentially just the total width $\Gamma$ of $d^*$) 
and a principal-value Green's function. The matrix element of the 
tensor-force operator is also different from Eq. (\ref{DD7D1}). 
The final result is rather similar to Eq. (\ref{DD7D1final}):

\begin{eqnarray}
\psi_3(p) \equiv \langle NN ^3D_3, p\vert d^* \rangle \approx
{\Delta E\over (\Delta E)^2 + (\Gamma/2)^2} h^*(p) \, ,
\label{nn3d3}
\end{eqnarray}
where $ \Delta E(p) = m^* - 2E_N(p) $,

\begin{eqnarray}
h^*(p) = \sqrt{12\over 5\pi^3} N_0^*{\rm e}^{-p^2/2\beta^{*2}}
\int v_{18}(q) {\rm e}^{-q^2/2\beta^{*2}} 
\nonumber \\ \times j_2(ix^*) q^2 dq \, ,
\label{hpstar}
\end{eqnarray}
and $x^* = pq/\beta^{*2}$.

  In Eq. (\ref{nn3d3}), a rapidly changing factor 
$ \Delta E/ ((\Delta E)^2 + (\Gamma/2)^2) $ has been separated 
from a function $h^*(p)$ that does not depend explicitly on 
$m^*$ or $\Gamma$. $ \Delta E$ changes sign as the nucleon 
momentum $p$ increases above that value $p_0$  at which 
$ \Delta E(p)$ vanishes. Above $p_0$, $ \Delta E$ is negative 
so that the perturbative wave function $\psi_3(p)$ tends to have 
the same phase relation to the driving $\Delta\Delta(^7S_3)$ 
wave function as $\psi_7(p)$ is to its own driving wave function. 
This means that diagrams \ref{figFeyn}b and c tend 
to interfere constructively at large momentum transfers, 
but destructively at small momentum transfers.

  The falloff parameter $\beta^*$ and 
normalization constant $N_0^*$ appearing in Eq. (\ref{hpstar})
are those for the single-Gaussian approximation to 
the $d^*(\Delta\Delta\,^7S_3)$ wave function.

  In the actual calculation, a three-term Gaussian fit to the 
two-centered $d^*$ wave function is used. The results for 
potential A28 are shown as long dashed curves in 
Fig. \ref{fig3d3wf} for $h^*(p)$ and the radial part 
$\psi_3(p)$ of Eq. (\ref{nn3d3}) 
calculated with $m^* = 2100$ MeV and $\Gamma = 100$ MeV. This 
spin triplet wave-function component has the perturbative 
normalization of $P_3 = 0.038$, or 3.8\%. This normalization is 
very sensitive to the choice of the decay width, increasing to 
0.55 if the width is decreased to 10 MeV. The 
rapid increase comes from the roughly $1/\Gamma$ behavior of 
the amplitude of the kink where the wave function oscillates 
rapidly from a positive value to a negative value as the 
momentum $p$ goes through the zero of $\Delta E$.

  The great sensitivity of the $^3D_3$ wave function to the 
width $\Gamma$ does not mean that the inelastic production 
amplitude increases just as dramatically. The wave function 
changes sign in the kink, leading to much cancellation in 
its contribution to the inelastic RME. Consequently, most RME's 
increase by only 25\% when the width decreases from 100 MeV to 
10 MeV, and approach stable values for smaller widths.

  Although the dependence on $\Gamma$ as described above is 
technically correct in a narrow sense, it is also 
counter-intuitive in that a smaller width with weaker coupling to 
the $NN$ channel somehow leads to a stronger $^3D_3$ component.
The truth is that in the discussion just given, the width 
is treated as if it were a free parameter when it actually is not. 
Rather once the dynamics is chosen, a certain width is implied, 
and must be used in Eqs. (\ref{DD7D1final}) and (\ref{nn3d3}). 

  Hence it is desirable to have a more consistent treatment. I 
shall first calculate the width for each input potential at 
each dibaryon mass $m^*$, and then use this calculated width 
to calculate admixed components.

\subsection{ Decay width of $d^*$ }

  The decay width of $d^*$ into the $NN$ ($\pi NN$, $d\pi \pi$) 
channel has been estimated in 
\cite{Wong98} (\cite{Wong98a}, \cite{Wong99}). It is clear that 
the $NN$ channel is by far its dominant decay channel; it is 
the only channel that has to be included in the present study. 
The $d^*$ decay widths into $NN$ for the A28 potential 
\cite{Wir84} and five different Bonn 
potentials \cite{Mac87,Mac89} are easily calculated by the same 
method. The results are shown in Table 1 for $m^* = 2100$ MeV and 
a single Gaussian wave function for $d^*$ of radius $r^* = 0.7$ fm. 

  The potentials are ordered in Table 1 in decreasing values of 
their deuteron $D$-state probability $P_{\rm D}$. The differences 
in $P_{\rm D}$ have been obtained by adjusting the amount of 
short-range repulsion in the tensor potential by changes in the 
coupling constants and baryon form factors.

  These potentials differ from one another in other interesting 
ways. Potential A28 has a $NN \leftrightarrow \Delta\Delta$ 
tensor potential from $\pi$ exchange with an overall strength 
$f_{\pi \Delta N}^2/f_{\pi NN}^2$ relative to the $NN$ tensor 
potential given the Chew-Low value of 4 \cite{Bro75}. The 
potential function has the Yukawa form corresponding to a meson propagator $1/\omega^2(q)$ in momentum space, where 
$\omega^2(q) = q^2 + m^2_{\rm meson}$. Its short-distance cutoff 
is specified in coordinate space using a function that bends the potential back to the origin to simulate the contribution of 
$\rho$ exchange \cite{Gre74}. 

  The Bonn potentials use nucleon form factors for the cutoff, 
with a monopole form for the $\pi$ vertex and usually a dipole 
form for the $\rho$ vertex. The cutoff masses $\Lambda_i$ used 
are shown in the table. The meson ($m$) coupling constants used 
here for the Bonn-A, B, 
and C potentials \cite{Mac89}, denoted here as potentials BA, BB 
and BC, and for the ``Full'' Bonn potential FB \cite{Mac87}, are 
the quark-model values based on the ratio 
$f_{m\Delta N}^2/f_{mNN}^2 = 72/25$ \cite{Bro75}. Potentials BA, 
BB and BC are relativistic momentum-space potnetials. FB is a relativistic model with energy-dependent meson propagators. The  
coupled-channel III, denoted here as CC3, is nonrelativistic 
and uses a strong-coupling ratio of about 4.4.

  The second numerical entry given in the first line for each 
potential is the width calculated with the $\pi$-exchange only of 
the potential. The result is clearly correlated with cutoff mass $\Lambda_\pi$ in the Bonn potentials. It decreases roughly 
monotonically from 130 MeV to only 30 MeV as we go from potential 
BC to BB, BA, Full Bonn (FB) 
and CC3 potentials as $\Lambda_\pi$ decreases from 3.0 to 0.8 GeV. 

When the $\rho$-exchange contribution is added, the resulting 
width, shown as the third numerical entry in the first line,
decreases dramatically to the range of 57 to 20 MeV. The trend 
appears to be correlated with the decreasing value of the 
deuteron D-state probability $P_{\rm D}$, except for the special 
case of potential FB discussed separately. It is 
of course the differences in $P_{\rm D}$ caused by different 
partial cancallations by the $\rho$-exchange potential that 
distinguish between these potentials in the first place.

  The result for the FB potential is different because the 
$NN \leftrightarrow \Delta\Delta$ potential generated by a simple 
quark-model prescription is not that shown in the first line of 
the results given in the table for the FB potential, but on the 
fourth line, where the width of 17 MeV is in rough agreement 
with the value for potential BA and with the trend dictated by 
$P_{\rm D}$. Note that the $\rho\Delta N$ form factor is a 
monopole, a difference noted in the table by the entry 
$[(n_\pi, n_\rho) = ]$ $(1,1)$ under the column ``Specials''. 
The results of line 4 for the FB potential are also the results 
reported previously in \cite{Wong98}.

  In the actual FB potential, the meson-$\Delta N$ vertices are 
further modified: 
First, the $\rho\Delta N$ form factor is changed from a monopole 
to a dipole form with no change in the cutoff mass $\Lambda_\rho$. 
This change of the form factor causes the $\rho$ contribtuion to 
be reduced significantly, thus causing the ``unexpected'' increase 
in the decay width.

  A second change made in the actual FB potential is to decrease 
the cutoff mass $\Lambda_\pi$ from 1.3 to 1.2 GeV. 
This change is dictated by the need to prevent the uncorrelated 
2$\pi$ contributions to the potential in the lower partial waves 
from becoming unmanageably large \cite{Mac87}. a situation that 
is not yet encountered in our lowest-order calculation. As far 
as our lowest-order result is concerned, Table 1 shows that the 
effect of this change is quite minor.

  The widths for potential CC3 also stand out from the trend. 
Comparing them with those for potential BB, which has a 
comparable $P_{\rm D}$, we can see that particularly the $\pi$  
contribution has been reduced significantly by the use 
of the much smaller cutoff mass, thus giving an abnormally small 
width in perturbation theory. Presumably, the channel-coupling 
potentials in CC3 will give rise to larger contributions from the 
long-range $\pi$-exchange potentials in higher orders. Hence it 
is likely that the large difference seen in the lowest-order 
calculation reported here will not persist so noticeably in 
higher orders.
 
  To summarize, the $d^*$ decay width appears to be controlled to 
a good extent by $P_{\rm D}$, and has a value in the range 15-60 
MeV. In comparison, the value of 100 MeV for potential A28 is more 
appropriate to $\pi$-exchange only. In other words, the 
simulation of $\rho$ exchange contained in it has not been 
very effectively in the long-distance part of the tensor potential 
that controls the decay width.

  I now turn to the many remaining entries in Table 1. The 
second line for each potential, or for each group within a 
potential, contains results obtained by replacing each cutoff 
factor $1/[1 + (q^2/\Lambda^2)]$ by the Gaussain 
$\exp [-(q^2/\Lambda^2)]$. This change, marked in the column ``Specials'' by the symbol G, improves the high-momentum 
behavior of various momentum integrals and also reduces the 
overestimate of the high-momentum wave function in our 
perturbative treatment. Its influence on our decay widths is 
minimal, except in potential CC3 where the small cutoff mass $\Lambda_\pi$ allows a relatively large effect. 

  In using the potential models discussed here, we either use 
the given coordinate-space potential $v(r)$ or the standard 
meson propagator $1/\omega^2(q)$, plus corrections from vertex 
form factors. The exception is the coupled-channel model CC3 
of \cite{Mac89}, for which we use the off-shell propagator

\begin{eqnarray}
{\cal P}(q) = \omega^{-1}(q) [\omega(q) + \Delta M]^{-1} \, ,
\label{propagator}
\end{eqnarray}
defined in the potential. Here $\Delta M = M_\Delta - M_N$. 

   The question could be raised as to whether one should use 
the off-shell propagator in these cross-channel potential as a 
general policy. Taken by itself, the answer is probably no. The 
reason is that 
the standard propagator leads to energy-independent Yukawa 
potentials that are actually superior to energy-dependent 
potentials arising from the use of off-shell propagators. It is 
not obvious, but the use of an energy-independent potential 
actually corresponds to the inclusion of that component of the 
two-baryon state having a virtual meson ``in the air'' \cite{Des88,Won94}. 

  Although the use of the off-shell propagators is not recommended, 
it is clear that their use will reduce the decay width calculated
with the standard propagators. To study the size of the reduction, 
I now repeat the calculation with off-shell propagators. The
results are shown on the third line of each group, and marked by the 
symbol $\Delta M$ under the ``Specials'' column. We see that this 
off-shell effect is quite large, especially on the 
$\pi$-exchange contributions. As a result, there is much closer cancellation than before between the $\pi$ and $\rho$ 
contributions, thus giving rise to a much smaller net decay width.

  In subsequent calculations in this paper, I shall use the A28 potentials to give an extreme case, and the BB/G potential for a 
more representative example, of the results expected for the 
electroproduction $T$ factors. The decay widths calculated for 
these two potentials are shown in Fig. \ref{figGamma} as functions 
of the dibaryon mass $m^*$. Gaussian vertex form factors are 
used in potential BB/G.
 
With these calculated widths, the $^3D_3$ component can now be 
estimated more reliably by perturbation theory. The results are 
given in Fig. \ref{fig3d3wf} as long dashed (solid) curves for
potential A28 (BB/G) using $m^* = 2100$ MeV and the calculated 
value of $\Gamma = 100$ MeV (38 MeV). Both the smooth part $h^*(p)$ 
and the complete radial wave function $\psi(NN\,^3D_3,p)$ of 
Eq. (\ref{nn3d3}) are shown. The perturbative normalization in 
this state for potential BB/G is $P_3 = 4.5\%$.

   The perturbative result for $\psi(\Delta\Delta\,^7D_1, p)$ of 
Eq. (\ref{DD7d1final}) calculated for potential BB/G with 
$m^* = 2100$ MeV and $\Gamma = 38$ MeV are also given in Fig. \ref{fig3d3wf} as a thick dashed curve when only the $\pi$-exchange 
part of the potential is included in the calculation, and as a 
thick solid curve when the $\rho$-exchange contribution is also included. The large reduction in the wave function caused by the inclusion of $\rho$-exchange is worthy of note. 
The perturbative normalization in this state for the complete 
BB/G potential is $P_7 = 0.38\%$.

\subsection{ Electroproduction $T$ factors }

  The inelastic production amplitudes from the dominant 
deuteron $NN(^3S_1)$ component to the $NN(^3D_3)$ component of 
$d^*$, as described by diagram \ref{figFeyn}c can finally be 
calculated. The major difference from the procedure described in  
Sec. III is that the radial wave function 
$\psi_3 \equiv d^*(NN\, ^3D_3)$ is now too complicated to be 
expanded readily in terms of harmonic-oscillator wave functions. 
It is kept in numerical form so that the final result, instead of 
being entirely analytic, now requires a one-dimensional numerical 
integration. 

   To simplify the calculation, the deuteron $d(^3S_1)$ wave 
function is expressed as a sum of three Gaussians. However,
results need to be given only for a single Gaussian. For 
this purpose, I use the notation of Eq. (\ref{Gaussian}), but with 
un-starred parameters. All production amplitudes then have the 
generic form 

\begin{eqnarray}
\langle d(^3S_1)\Vert {\hat {\cal O}}(q) \Vert d_3^*\rangle
= c_0 N_0 {\rm e}^{-(q/2)^2/2\beta^2} {\rm e}^{-\alpha_0 q^2/2}
\nonumber \\ \times \int {\rm e}^{-p^2/2\beta^2} 
\psi_3(p) f_{\cal O}(p,q) p^2 dp \, ,
\label{rme1c}
\end{eqnarray}
but different integrand functions  

\begin{eqnarray}
f_{\cal O}(p,q) & = & j_2, \quad
 {\rm for} \: {\hat M_2}\, ; \nonumber \\
& = & {\sqrt{6}\over M_N} {2\beta^2\over q} j_2, \quad
{\rm for} \: {\hat T_{\rm c2}^{\rm el}}\, ; \nonumber \\
& = & -{\mu_{\rm s}(N)\over M_N} {c_1\over \sqrt{6}}
\left [ \left ({q\over 4} + {4\beta^2\over q} \right ) 
j_2 - ipj_1 \right], \nonumber \\
& &  {\rm for} \: {\hat T_{\mu 2}^{\rm el}}\, ; \nonumber \\
& = & {\mu_{\rm s}(N)\over M_N} {2\over \sqrt{21}}
\left [ \left ({q\over 4} + {10\beta^2\over q} \right )j_2 
- ipj_1 \right], \nonumber \\
& &  {\rm for} \: {\hat T_{\mu 3}^{\rm mag}}\, . 
\label{ffn1c}
\end{eqnarray}
Here $j_i \equiv j_i(ipq/\beta^2)$ is a spherical Bessel function, 
and $c_0 = \sqrt{7}$, $c_1 = 1$. 

  The production amplitudes from diagram 1a have a similar form,
but with $c_0 = \sqrt{3}$, $c_1 = 4$, and of course different 
radial wave functions. The baryon magnetic-moment parameter that 
appears is now $\mu_{\rm s}(\Delta)$ instead of $\mu_{\rm s}(N)$. 
I have verified explicitly that they give the 
same numerical values as the oscillator expressions given in Sec. 
III. The difference in the numerical coefficients $c_\alpha$ comes 
from the recoupling of angular momentum. These alone favor the amplitudes form diagram \ref{figFeyn}c by a factor of roughly 
$\sqrt{7/3}$, which is a ratio of two 6j-symbols.

  To perform the calculation for potential BB/G, I use a 
three-term approximation to the BB deuteron $NN(^3S_1)$ 
wave function \cite{Mac89}. The resulting perturbative 
deuteron $\Delta\Delta(^7D_1)$ wave function calculated from 
Eq. (\ref{DD7D1final}) is fitted to a 4-term harmonic-oscillator 
form. Both fitted wave functions are given in the Appendix. This 
fitted $\Delta\Delta(^7D_1)$ wave function is the one shown as a 
thick solid curve in Fig. \ref{fig7d1wf}.

   The calculated effective $T$ factors are shown in Fig. 
\ref{figEffT} with the curves defined in the legion.

  we can see that electroproduction RMS's from diagram 
\ref{figFeyn}a (in the case of A28) or b (BB/G) and diagram \ref{figFeyn}c interfere destructively at small $q$ values. At 
$q$ = 300 MeV/$c$, for example, the production amplitude that
 contributes to the Coulomb factor $T^L$ from diagram \ref{figFeyn}c 
for potential A28 is opposite in sign and about 3.6 times larger 
than that from diagram \ref{figFeyn}a. Of the increase, a factor $\sqrt{7/3}$ comes from a 6j-symbol, while the remaining factor 
of 2.4 comes from the wave functions in a radial integral. 
Hence $T^L$ is about 7 times that for diagram \ref{figFeyn}a alone. 

   The production amplitude from diagram \ref{figFeyn}c varies more 
strongly with the momentum transfer because of the much stronger momentum dependence of the $d^*(NN \,^3D_3)$ wave function it 
contains. This production amplitude changes sign just below 
$q = 1000$ MeV/$c$ where the curve for diagrams \ref{figFeyn}a and c
crosses that for diagram \ref{figFeyn}a alone the second time. This 
is above the interference zero that can be seen in 
Fig. \ref{figEffT}. After the sign change, the amplitudes from 
the two diagrams add constructively.

In the effective transverse $T$-factors for potential A28, 
the production amplitudes from the two diagrams are of comparable magnitudes but opposite signs at low momentum 
transfers. The resulting destructive interference is very severe, 
causing the total contribution to $T_{\rm eff}$ to be some two 
orders of magnitude smaller than that from diagram \ref{figFeyn}a 
alone. The interference eventually becomes constructive at roughly 
the same momentum transfer as for the Coulomb $T^L$ factor. 

The results for the BB/G potential is qualitatively similar, but 
the Coulomb $T^L$ factor is smaller at the smaller momentum transfers.

\section{ Results and discussions }

For all electron energies greater than a few GeV, the effective 
$T$ factors are essentially energy independent. The remaining 
factors in the integrated differential production cross section 
$d\sigma/d\Omega$ of Eq. (\ref{eq:dcs}) do depend on the 
energy, however. The differential cross section 
then increases roughly as the square of the scattered electron 
energy at comparable four-momentum transfers and small lab angles. 
This means that the cross section tends to increase quadratically 
with the incident electron energy as the latter increases.

The resulting angular distributions are shown in Fig. 
\ref{figAngDist} for 1 GeV electrons. The result for potential 
A28 is for diagrams \ref{figFeyn} a and c, while those for 
potential BB/G are for diagrams \ref{figFeyn}b and c. The 
interference between the two diagrams can be visualized more 
readily by also examining Fig. \ref{figAD1c} which show 
additional results for potential BB/G from the diagram 
\ref{figFeyn}c alone. 

  For the dibaryon mass $m^* = 2.1$ GeV, the sharp local minima 
near $60^\circ$ comes from the amplitude zero in the Coulomb 
$T^L$ factor. When diagram \ref{figFeyn}c appears alone, this zero 
comes from the sign change in the $d^*(NN\, ^3D_3)$ wave function. 
When both diagrams \ref{figFeyn}b and s are included, it comes 
from an interference 
zero in the sum of their amplitudes. The two diagrams interefere destructively below the sharp minimum, and constructively above it. 
For example, the integrated cross section at the lab angle of 
$30^\circ$ is 0.05 nb/sr for potential BB/G, but 0.11 nb/sr for 
diagram \ref{figFeyn}c alone.

  The over-estimate of the cross section by potential A28 can be 
seen in Fig. \ref{figAngDist}. At the lab angle of $30^\circ$, the 
cross section of 0.09 nb/sr for potential A28 is about twice as 
large as the value for potential BB/G. Other than this, both cross sections show the same decreasing trend with increasing lab angle.

   The Mott cross section $\sigma_{\rm M}$ and the target recoil 
factor $f_{\rm rec}$ 
do not depend on the dibaryon mass $m^*$. However, the magnitude 
of four-momentum transfer decreases as $m^*$ increases, while that 
of the three-momentum transfer increases. This leads to a rapid 
decrease in the kinetimatical factor $v_{\rm L}$. 

  The dynamical factor $T_{\rm eff}$ also changes with $m^*$. This 
is partly because of the three-momentum transfer appearing in it 
and partly because the $d^*$ wave function $d^*(NN\, ^3D_3)$ 
itself also changes. The interference between the two  
diagrams becomes rather complex. The Coulomb $T^L$ factor 
decreases rapidly with increasing $m^*$. At $m^* = 2.3$ GeV, the 
minimum in the angular distribution at $135^\circ$ is caused 
by the interference zero in both $T^L$ and $T^T$. At $m^* = 2.4$ 
GeV, the shallow minimum near $90^\circ$, comes from the 
combined effects of an interference zero in $T^T$ at $100^\circ$ 
and an interference zero in $T^L$ at $50^\circ$. The destructive interference between the two production amplitudes at small 
momentum transfers tends to become more severe as $m^*$ increases.

  Fig. \ref{figAngDist} shows how rapidly the cross section 
decreases with increasing $m^*$. At the lab angle of $30^\circ$, 
the integrated cross section with both diagrams present, which at 
0.05 nb/sr is already small for $m^* = 2.1$ GeV, now falls down to 
only 0.0005 (0.00006) nb/sr for $m^* = 2.3$ (2.4) GeV, as shown in 
Fig. \ref{figAngDist}. 

  The very small calculated cross sections obtained for both  
larger angles and larger $m^*$ suggest that the present
lowest-order picture might not be adequate under these 
circumstances. A much more elaborate calculation that includes 
higher-order diagrams as well as additional production 
processes important at these larger three-momentum transfers 
might have to be considered. 

The result reported here can be compared with a recent calculation 
by QSW \cite{Qing99}, who include diagram \ref{figFeyn}c and a 
production amplitude going to the $NN^*(1520)$ component of $d^*$. 
They find that diagram \ref{figFeyn}c dominates the calculated 
cross section for 1 GeV electrons scattered to a lab angle of 
$30^\circ$, and obtain a preliminary value for the integrated 
cross section greater than 10 nb/sr there. 

In order to compare with QSW, we show in Fig. \ref{figAD1c} our 
result for diagram 1c alone at the same $m^* = 2.1$ GeV but 
calculated with potential BB/G. Our cross section is only 0.11 
nb/sr at 30$^\circ$, two orders of magnitude smaller. 

One difference between the two calculations is that the 
baryon-baryon interaction used by QSW comes from \cite{Gom94} 
and contains only the $\pi$-exchange contribution. In potential 
A28 and certainly in the Bonn potentials used here, the 
baryon-baryon tensor potentials are significantly reduced by
cancellation against additional $\rho$-exchange contributions. 

To isolate the contribution of $\pi$-exchange alone, we repeat 
the calculation for diagram 1c alone with only one modification --- using the interaction of \cite{Gom94} (called here potential GO) 
to generate the perturbative $d^*(NN\, ^3D_3)$ wave function. 
The potential GO has the coupling constant $f_{\pi\Delta N}^2 = 
0.36$ ($ = f^{*2}/4\pi$ in the notation of \cite{Gom94}), roughly consistent with those shown in Table \ref{tab-width}.
A monopole form factor for the $\pi\Delta N$ vertex is used with 
the cutoff parameter $\Lambda_\pi = 1250$ MeV. The $^3S_1$ wave
function used in this calculation remains that of the Bonn B 
deuteron. This is not very different from the Bonn C wave function 
used by QSW.

  The resulting cross section, for $m^* = 2.1$ GeV, is shown in 
Fig. \ref{figAD1c} as a dotted curve. The calculated cross
section at 30$^\circ$ is 0.41 nb/sr, almost four times bigger than 
the BB/G result. This shows that the neglected $\rho$-exchange contributions do reduce the cross section significantly. 

The situation remains qualitatively the same when diagram 
1a is also included: The cross section for potential GO at 
30$^\circ$ is 0.29 nb/sr, a factor of 6 greater than the value 
of 0.05 nb/sr for potential BB/G.

The cross section for diagram 1c alone for potential BB/G 
calculated with $m^* = $ 2.3 and 2.4 GeV are also given in the 
figure and compared with the results for $m^* = $ 2.1 GeV 
with only diagram 1c and with both diagrams present.

After accounting for the difference in dynamical inputs, I find 
a remaining discrepancy of more than an order of magnitude 
between my calculation and that of QSW. This remaining difference 
must be due to differences in the calculational methods used. The 
most important difference seems to be my use of a principal-value
 Green's function in my $d^*(NN\,^3D_3)$ wave 
function compared to the use of an outgoing-wave boundary 
condition in QSW. The QSW calculation thus gives an additional 
term in the production amplitude from diagram \ref{figFeyn}c 
that contains an energy-conserving $\delta$-function. This 
additional term has a part that is the decay amplitude of $d^*(\Delta\Delta$ into the $NN$ channel. I do not include this 
part as the physical electroproduction production amplitude.
Additional studies must be made to understand 
if this could account for all the remaining disagreement.

Returning to the general problem of calculating the 
electroproduction cross section of $d^*$, there are of course the additional uncertainties in the predicted mass and structure of 
$d^*$ itself. Given the experimental non-observation of $I$=0 
dibaryons in the mass range 2.00-2.23 GeV \cite{Lis82}, future 
experimental searches and theoretical studies should probably 
move to higher masses. For electroproduction, the rapid decrease 
of the cross section with increasing $d^*$ mass is a cause of 
concern if it persists in higher-order calculations.

The theoretical uncertainty in the $d^*$ structure remains a 
major obstacle. The didelta model of $d^*$ used here 
is very crude, and may not be adequate at short distances where 
the baryons overlap. Other exotic components in the deuteron 
wave function must also be taken into consideration, especially at 
the larger momentum transfers. The possibility of quark
delocalization and color screening remains an open question, but 
the failure to see any $d^*$ dibaryon in the experiment of 
\cite{Lis82} is discouraging.

All these issues make it clear that much more work remains to be 
done before a quantitative description of the $d^*$ production 
cross section can be achieved, given any theoretical model for 
$d^*$. The preliminary study given in this paper does suggest that 
the electroproduction cross sections to $d^*$ are likely to be 
very small.

Perhaps more interestingly, the existence of $d^*$ at low masses 
can now be viewed from a rather elementary perspective. Consider 
a very naive model of $d^*$ containing a probability 
$P_{\Delta\Delta}(d^*)$ of the $\Delta\Delta(^7S_3)$ configuration 
and the remaining probability, $1 - P_{\Delta\Delta}(d^*)$, of a 
closed-channel configuration that does not decay at all into any 
channel. The two pieces of information: 

\begin{enumerate}
\item[(a)]
the best available experimental upper bound of 0.08 MeV obtained 
by \cite{Lis82} for its decay width into the $NN$ channel near 
2.1 GeV, and 
\item[(b)]
the best theoretical estimate of its width of 
about 40 MeV given in this paper if $d^*$ is a didelta,
\end{enumerate}
if taken literally, would require that $P_{\Delta\Delta}(d^*)$ 
cannot exceed something of the order of $0.2\%$. This seems to 
suggest that any model of $d^*$ containing a much larger 
$P_{\Delta\Delta}(d^*)$ might already have been excluded by the 
measurements of \cite{Lis82}.

I wish to thank Fan Wang, Terry Goldman, Stan Yen, Earle Lomon, 
Bob Wiringa, Faqir Khanna and R. Dymarz for many stimulating conversations and correspondence. I am indebted to Stan Yen for 
asking about the resonance contribution to the total cross 
section measurement of Lisowski {\it et al.} \cite{Lis82} that 
led to the experimental upper bound given in the Introduction.

\appendix
\section{Wave functions used in the calculation}

Many of the wave functions used have been fitted to the form 

\begin{equation}
\psi (p) \approx  \sum_{i=1}^3 c_i \psi_i(p),
\label{fittedwf}
\end{equation}
where $\psi_i(p)$ is a normalized harmonic-oscillator wave 
functions (HOWF) like that shown in Eq. (\ref{d7d1-wf}), with 
the falloff parameter

\begin{equation}
\mbox{\boldmath $\beta$}^2 =  (\beta_1^2, \beta_2^2, ...) \, .
\label{betasq}
\end{equation}
The dimensionless expansion coefficients are

\begin{equation}
{\bf c} = (c_1, c_2, ...)\, . 
\label{amp}
\end{equation}

In order to emphasize the stronger high-momentum components in 
these short-distance wave functions, the range parameters are 
obtained by minimizing the {\it percentage} mean-square deviation. 

The fitted parameters are given in Table \ref{tab-wf}. Each fitted 
wave function has been renormalized in order to change the raw 
percent probability to the value the original wave function has, 
as indicated in the third line of the table for each state.

\begin{figure}
\caption{Electroproduction of $d^*$ from the deuteron: 
(a) from the deuteron $\Delta\Delta \, ^7D_1$ state, 
(b) from the perturbative deuteron $\Delta\Delta \, ^7D_1$ 
wave function, and 
(c) from the deuteron $^3S_1$ state to the perturbative  
$d^*(NN \, ^3D_3)$ wave function.}
\label{figFeyn}
\end{figure}

\begin{figure}
\caption{The effective T factors appearing in the differential 
electroproduction cross section of $d^*$ from the deuteron at 
different electron lab energies as calculated from 
diagram 1a for potential A28.}
\label{figEffT1a}
\end{figure}

\begin{figure}
\caption{Comparison of the perturbative wave functions 
$\psi_7(p)$ in momentum space of the deuteron 
$\Delta\Delta \, ^7D_1$ state for the potentials A28 and BB/G 
to the exact value for the A28 potential. Here $\mu = 138.0$ MeV
is the pion mass.}
\label{fig7d1wf}
\end{figure}

\begin{figure}
\caption{The perturbative wave functions of the $d^*$($NN \,
^3D_3$) state for the potentials A28 and BB/G showing
(a) the smooth function $h^*(p)$, and (b) the wave function 
$\psi_3(p)$ in momentum space. 
Here $\mu = 138.0$ MeV is the pion mass.}
\label{fig3d3wf}
\end{figure}

\begin{figure}
\caption{The decay width $\Gamma$ of $d^*$ as a function of 
the $d^*$ mass $m^*$.}
\label{figGamma}
\end{figure}

\begin{figure}
\caption{Comparison of the effective T factors in the 
electroproduction cross section of $d^*$ from the deuteron 
as calculated from diagrams 1a and c for potential A28 and 
from diagrams 1b and c for potential BB/G at the electron 
lab energy of 27.5 GeV with the results for diagram 
1a alone for potential A28.}
\label{figEffT}
\end{figure}

\begin{figure}
\caption{Integrated differential cross sections $d\sigma/d\Omega$
in the lab for the production of $d^*$ from deuteron for 1 GeV 
electrons calculated for potentials A28 and BB/G at different 
dibaryon masses $m^*$.}
\label{figAngDist}
\end{figure}

\begin{figure}
\caption{Integrated differential cross sections $d\sigma/d\Omega$
in the lab for the production of $d^*$ from deuteron for 1 GeV 
electrons calculated for potential BB/G at different dibaryon 
masses $m^*$ when only diagram 1c is included. The dotted curve
gives the result when the perturbative $d^*(NN\,^3D_3)$ wave 
function is calculated using the potential GO instead. 
For comparison, the result for $m^* = 2.1$ GeV when both diagrams 
1b and c are included is shown as a thick solid curve.}
\label{figAD1c}
\end{figure}

\begin{table}
\caption{Decay width $\Gamma(d^*)$ in MeV in the didelta model of 
the $d^*$ for different baryon-baryon interactions. }
\begin{tabular}{cccccccc}
Potential& $P_{\rm D}(\%)$& $\pi$ only &  $\pi + \rho$& $f_{\pi\Delta N}^2{}^a$& $\Lambda_\pi{}^b$ & $\Lambda_\rho{}^b$ &  Specials   \\  
\hline
A28& 6.13& 100&    &  0.324&  &  &  \\
\hline
BC&  5.61& 133&  57&  0.221&  3.0&  1.7&   \\
&        & 132&  58&       &     &     &  G  \\
&        &  56&  21&       &     &     &  G/$\Delta$M  \\
\hline
BB&  4.99& 101&  38&  0.224&  1.7&  1.85&  \\
&        &  98&  38&       &     &      &  G  \\
&        &  41&  12&       &     &      &  G/$\Delta$M \\
\hline
CC3& 4.87&  30&  20&   0.35&  0.8&  1.35&  $\Delta M$ \\
&        &  21&  13&       &     &      &  G/$\Delta$M \\
\hline
BA&  4.38&  79&  17&  0.229&  1.3&  1.95  \\
&        &  74&  15&       &     &     &  G \\
&        &  31&   4&       &     &     &  G/$\Delta$M \\
\hline
FB&  4.25&  68&  44&  0.224&  1.2&  1.4&  \\
&        &  62&  40&       &     &     &  G \\
&        &  26&  15&       &     &     &  G/$\Delta$M \\
\hline
&        &  76&  17&       &  1.3&  1.4& (1,1)   \\
&        &  71&  15&       &     &     &  G \\
&        &  29&   4&       &     &     &  G/$\Delta$M

\end{tabular}
$^a$  dimensionless \\
$^b$  in GeV
\label{tab-width}
\end{table}

\begin{table}
\caption{Wave functions used in the calculation.}
\begin{tabular}{cc}
Wave function & Parameters   \\  
\hline
$d^*(NN\, ^7S_3)$ &
$\mbox{\boldmath $\beta$}^2 =$ 2.0753 (1, 1.25, 1.46) fm$^{-2}$\\
& {\bf c} = (14.8312, -27.4124, 13.3505) \\
& $100.19\% \rightarrow 100.00\%$  \\
\hline
$d(NN\, ^3S_1)$ A28&
$\mbox{\boldmath $\beta$}^2 =$ 0.04056 (1, 7.25, 169) \\
& {\bf c} = (0.5250, 0.5900, -0.0573) \\
& $93.17\% \rightarrow 93.31\%$  \\
\hline
$d(NN\, ^3S_1)$ BB &
$\mbox{\boldmath $\beta$}^2 =$ 0.02298 (1, 5.15, 21.9) \\
& {\bf c} = (0.3192, 0.4859, 0.3486) \\
& $95.04\% \rightarrow 95.01\%$  \\
\hline
$d^*(\Delta\Delta\, ^7D_1)$ A28&
$\mbox{\boldmath $\beta$}^2 =$ 0.1769 (1, 5.0, 15.0) \\
& {\bf c} = (0.00118, 0.01537, 0.05407) \\
& $0.4182\% \rightarrow 0.4190\%$  \\
\hline
$d^*(\Delta\Delta\, ^7D_1)$ BB/G&
$\mbox{\boldmath $\beta$}^2 =$ 0.9395 (1, 3.18, 8.26, 28.0) \\
& {\bf c} = (0.00316, 0.02245, 0.02422, 0.02955) \\
& $0.379\% \rightarrow 0.383\%$  

\end{tabular}
\label{tab-wf}
\end{table}

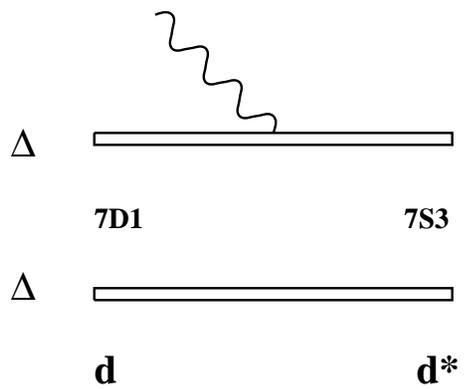
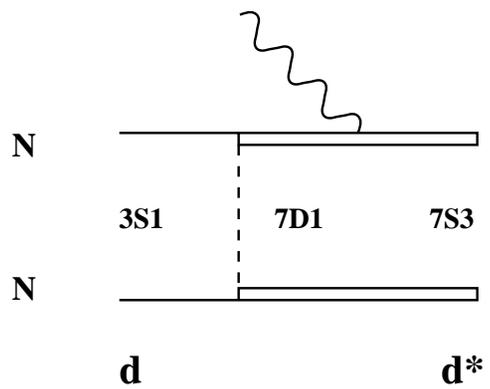
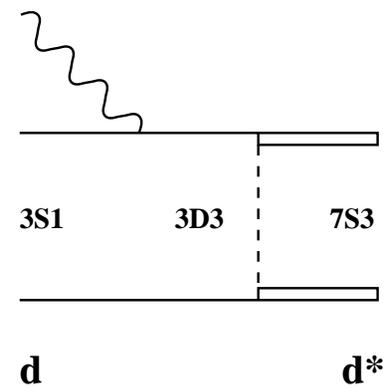
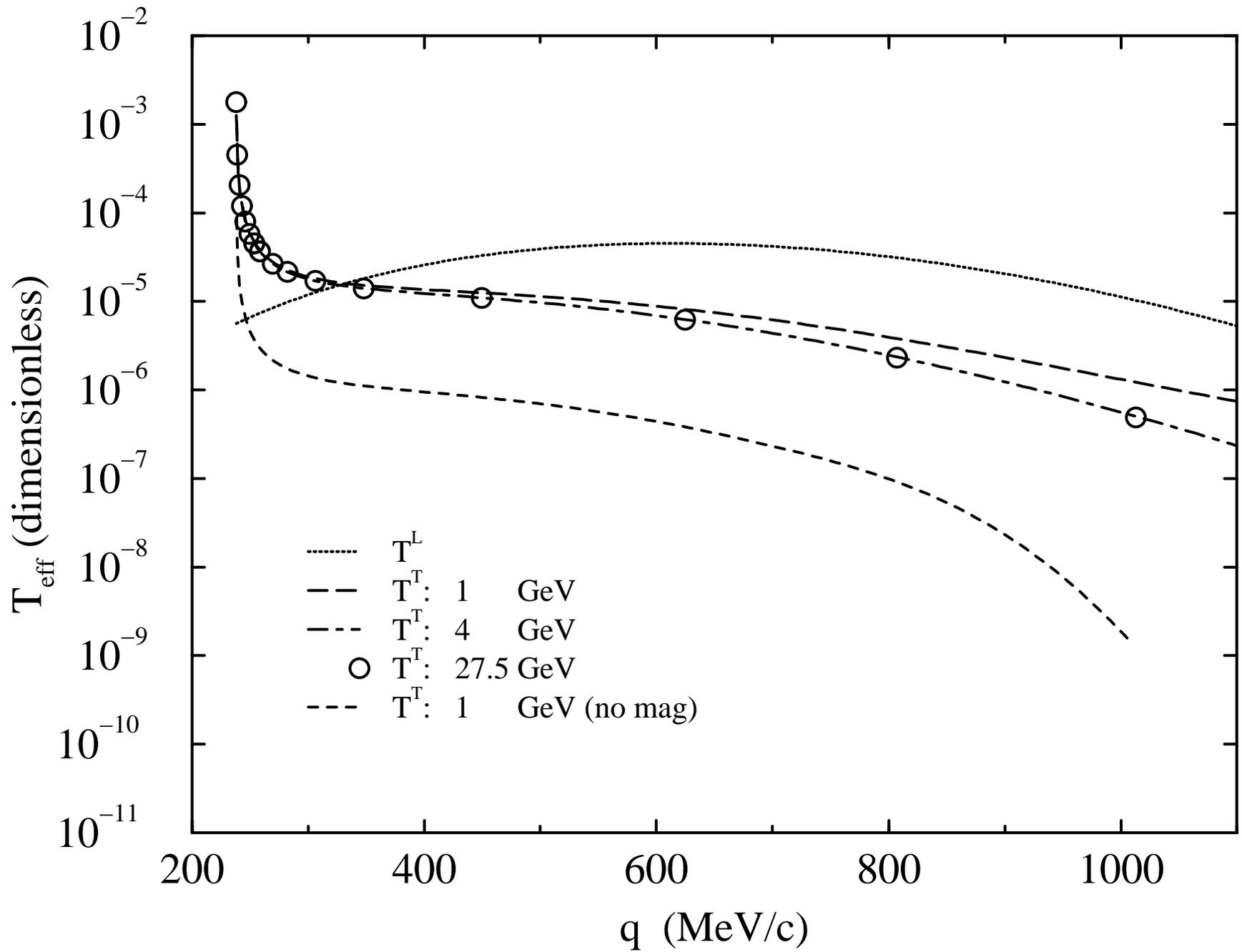
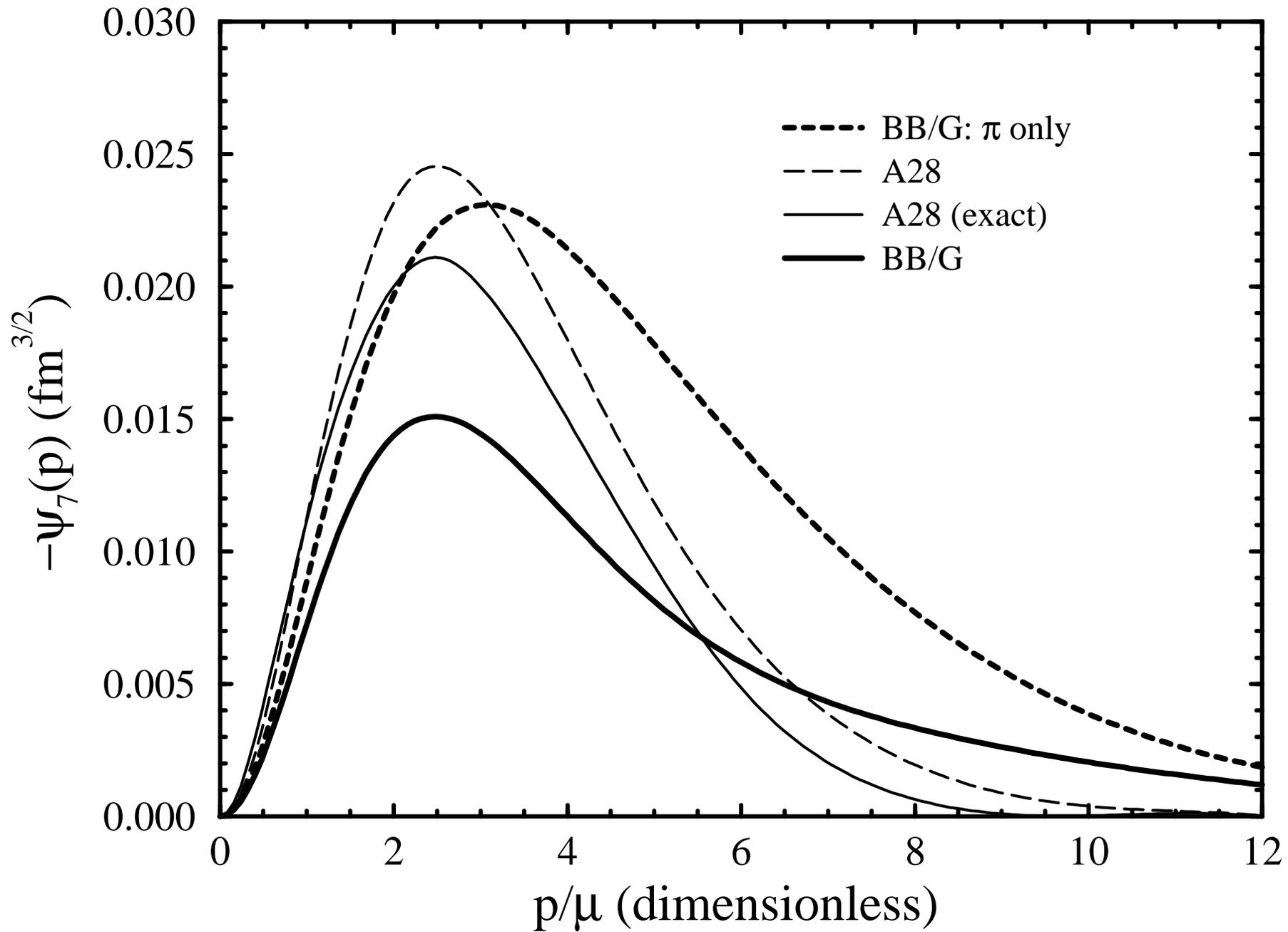
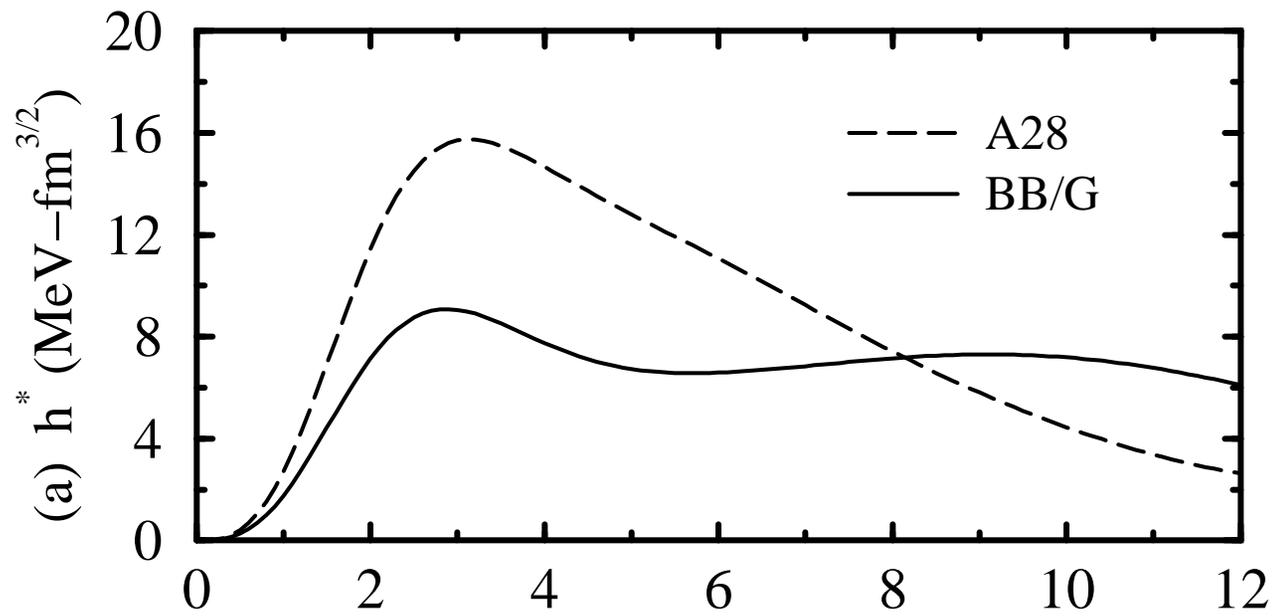
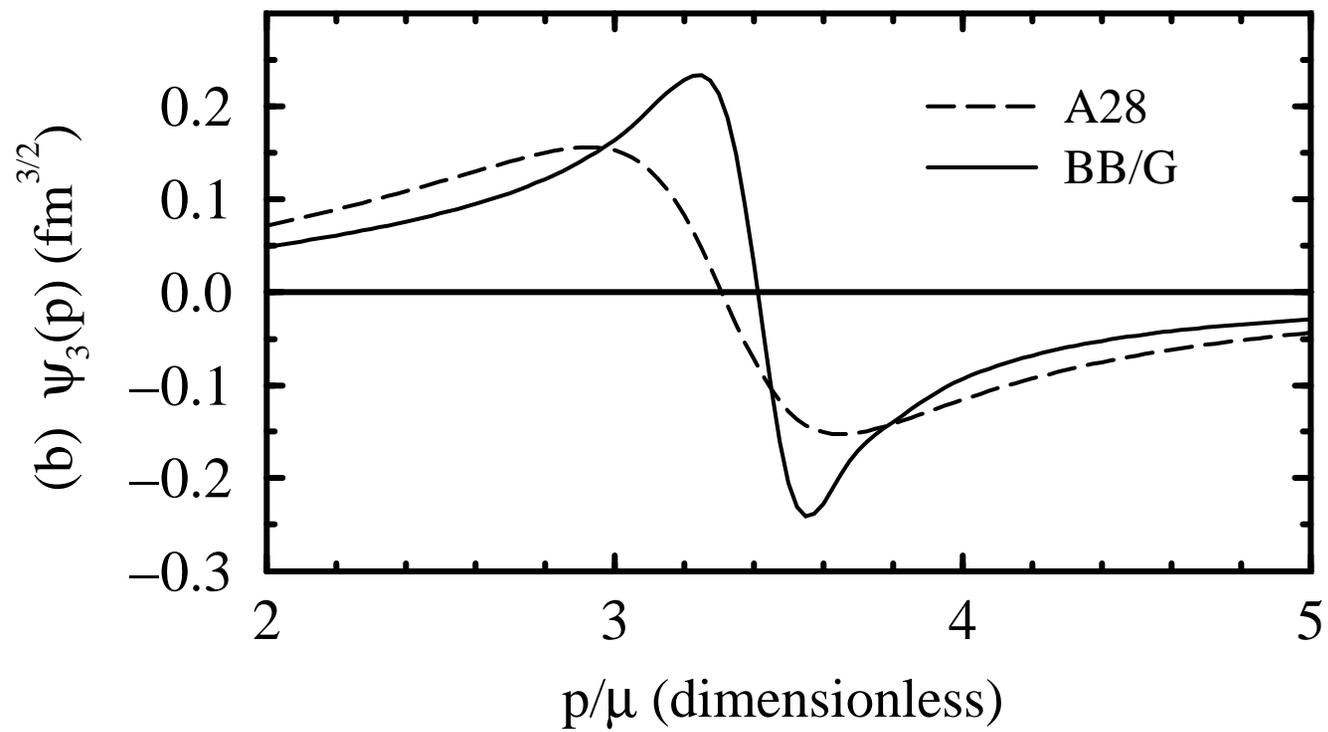
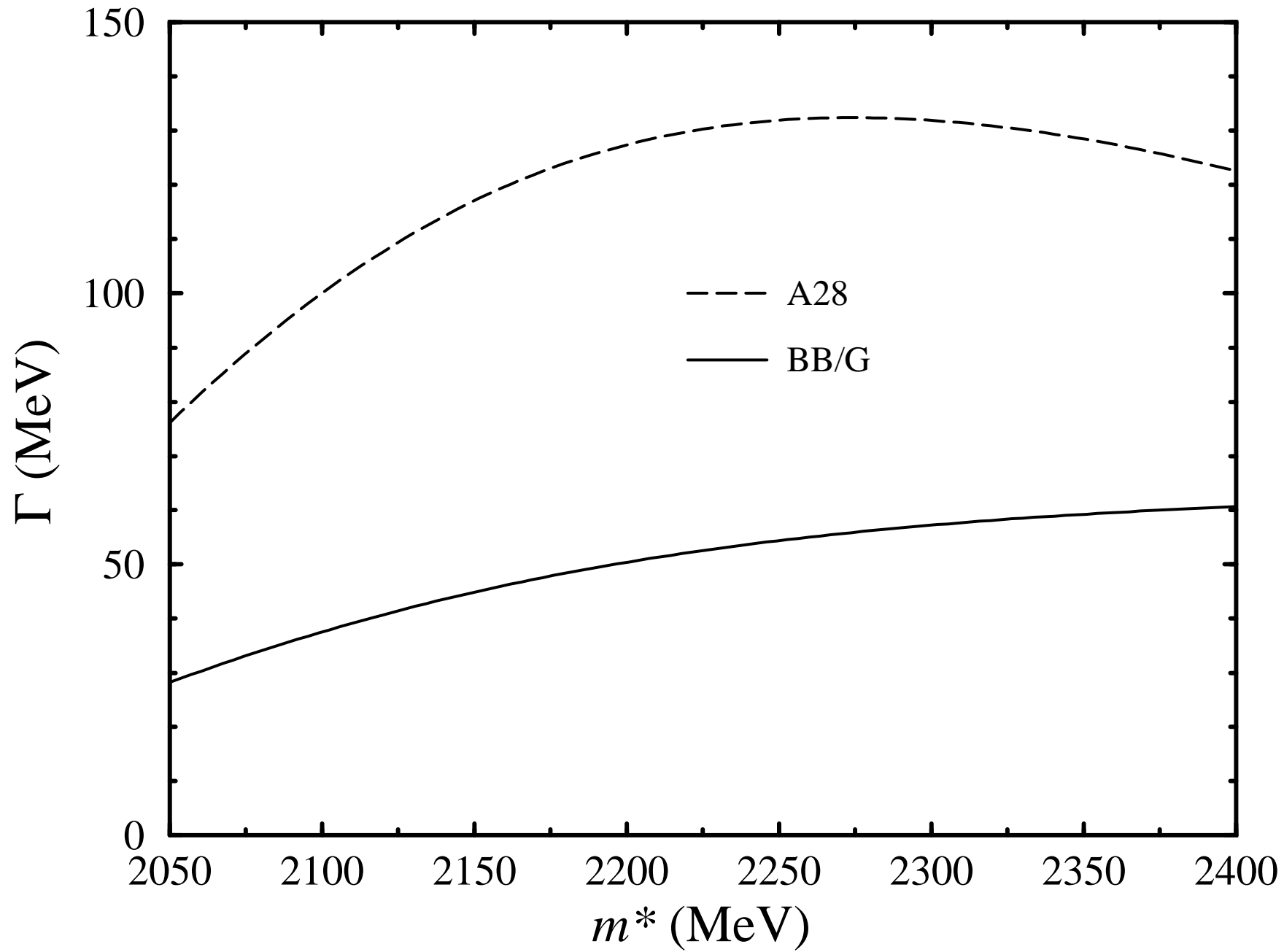
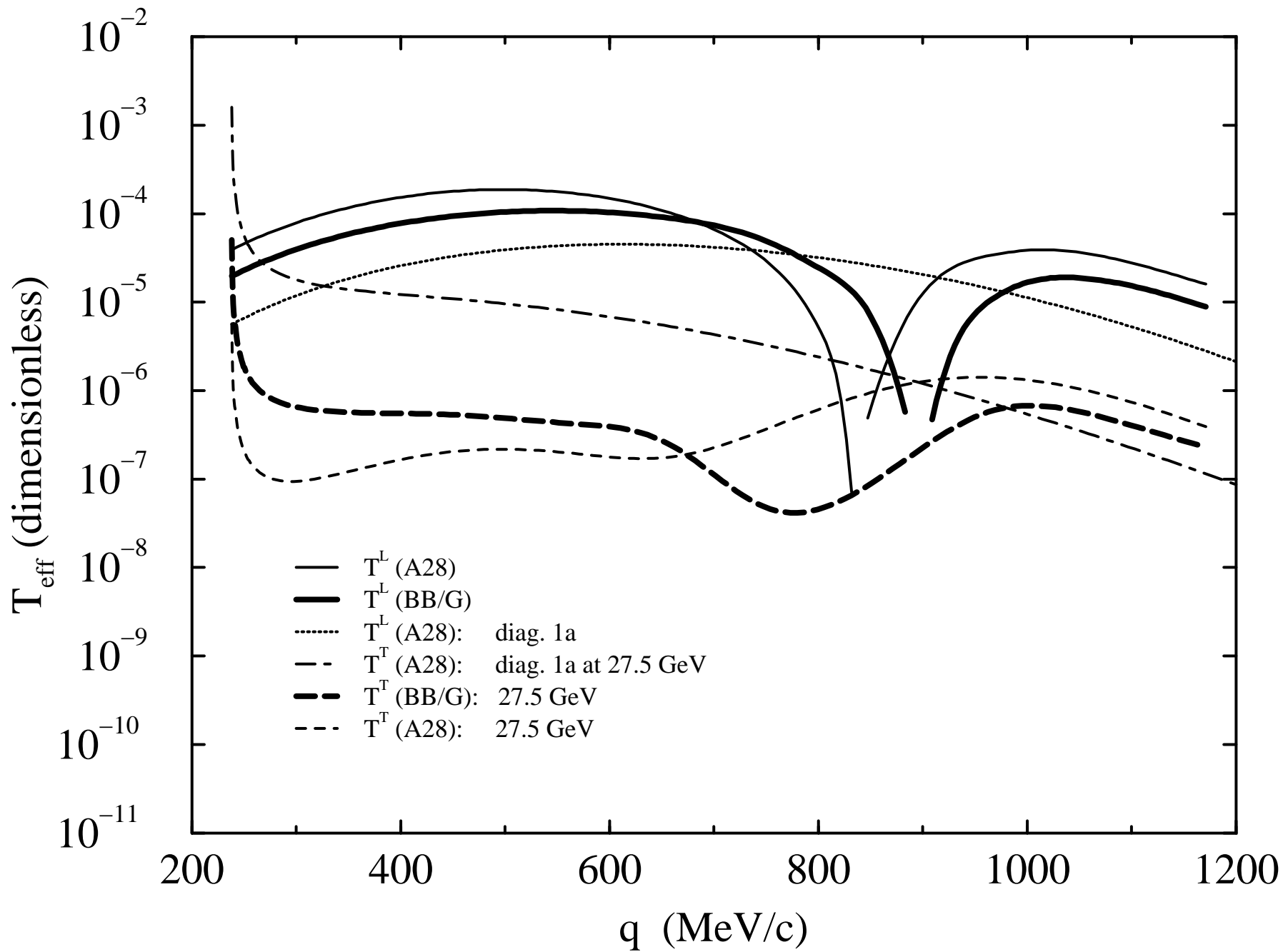
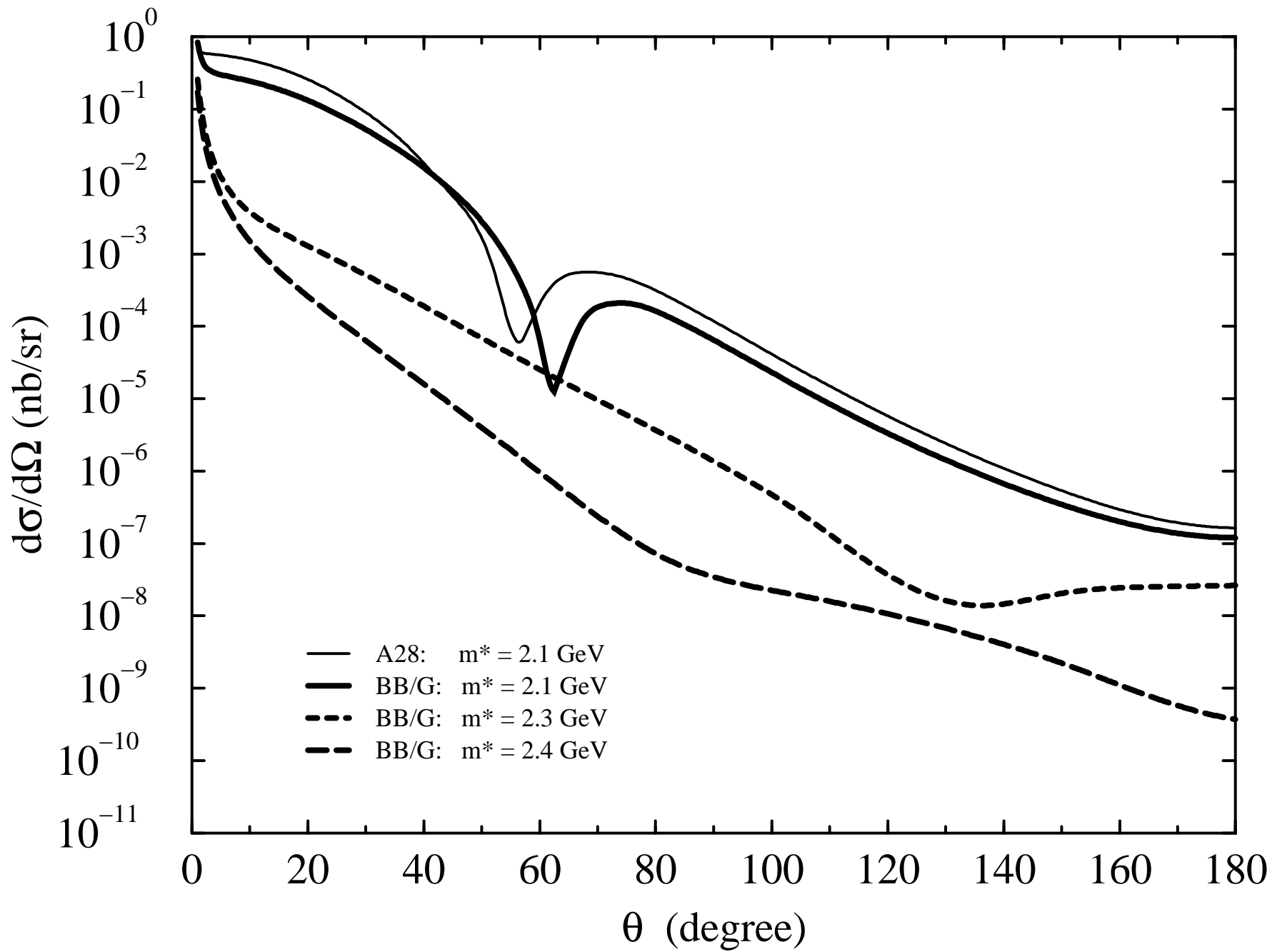
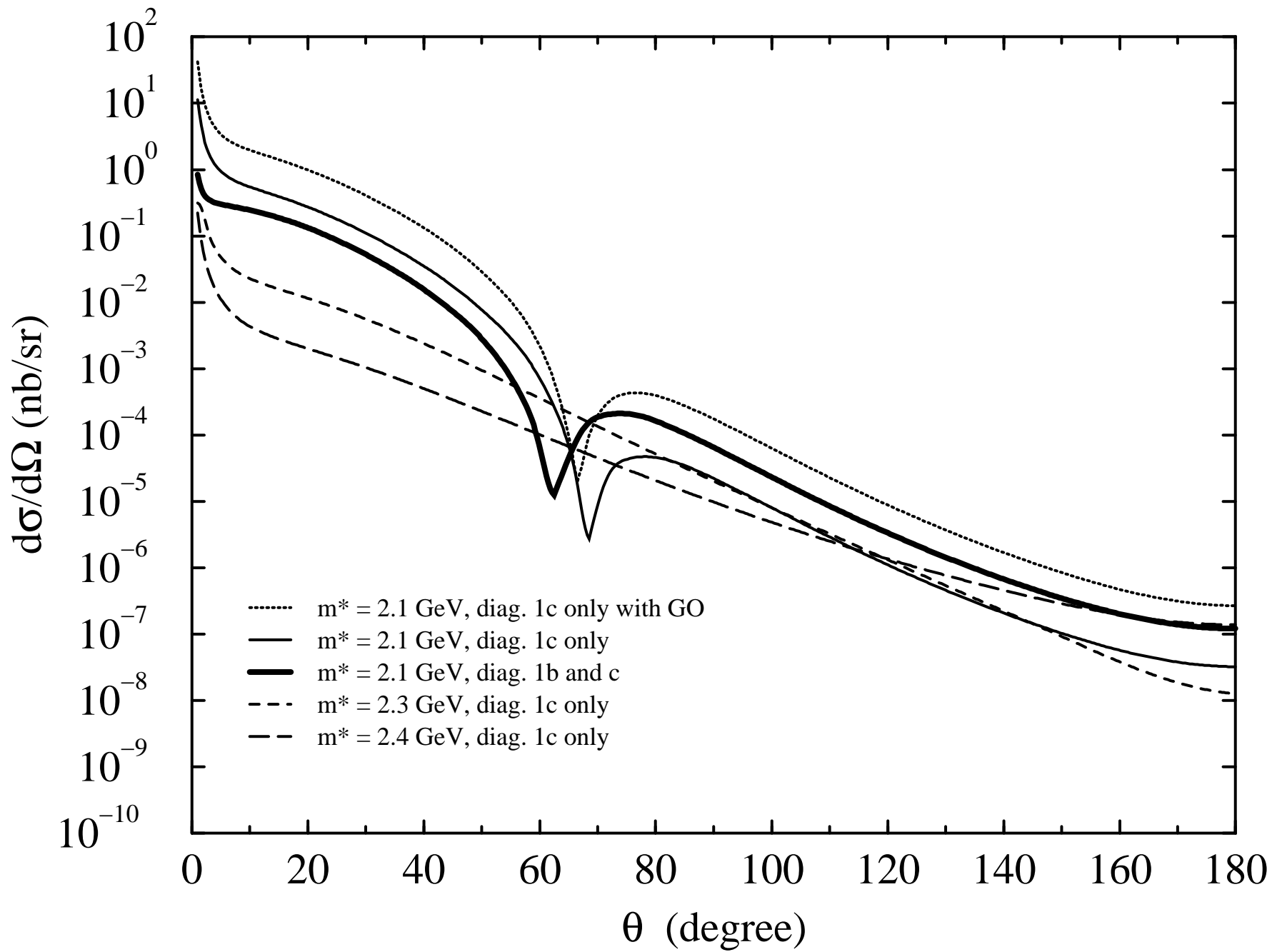
\end{document}